\documentclass[a4]{article}

\def\l{{\lambda}}

\def\a{{\alpha}}
\def\b{{\beta}}
\def\g{{\gamma}}

\def\C{{\cal C}}
\def\L{{\cal L}}

\def\te#1{{\widetilde{#1}}}

\def\on#1#2{\mathop{\vbox{\ialign{##\crcr\noalign{\kern2pt}
$\scriptstyle{#2}$\crcr\noalign{\kern2pt\nointerlineskip}
\kern-2pt$\hfil\displaystyle{#1}\hfil$\crcr}}}\limits}
\def\nn{ \nonumber }
\def\bq{ \begin{equation} }
\def\eq{ \end{equation} }
\def\ben{ \begin{eqnarray} }
\def\en{ \end{eqnarray} }
\def\ll{ \label }
\def\frac#1#2{{#1\over #2}}
\def\dfrac#1#2{{\displaystyle{#1\over#2}}}

\begin{document}
\title{The Lax pairs for the Holt system.}
\author{
A.V. Tsiganov\\
\it\small Department of Mathematical and Computational Physics,
 Institute of Physics,\\
\it\small  St.Petersburg University,
 198 904, St.Petersburg, Russia.\\
\it\small e-mail: tsiganov@mph.phys.spbu.ru}
 \date{}
\maketitle

\begin{abstract}
By using known non-canonical transformation between the Holt system and
the Henon-Heiles system the Lax pairs for all the integrable cases of the
Holt system are constructed from the known Lax representations for the
Henon-Heiles system.
\end{abstract}


\section{Introduction}
\setcounter{equation}{0}
The Holt system is defined by the Hamilton function
\bq
H=\dfrac12\,(\,p_X^2+p_Y^2\,)+\a\,X^{-2/3}\,\,
(\,\dfrac{3\,\b}4\,X^2+Y^2+\g\,)\,.
\ll{holt}
\eq
Only three integrable cases are known \cite{ho82,rgb89}
\bq
(i)~\b=1\,,\qquad(ii)~\b=6\,,\qquad(iii)~\b=16\,,
\eq
while the remaining parameters $\a$ and $\g$ be an arbitrary constants.
These parameters were isolated by the singular analysis \cite{rgb89},
although the second integrals may be obtained directly \cite{ho82,hi87}.

By integrability here we mean existence of a second independent integral
of motion $K$, and in this case the Liouville theorem implies that the
problem can be solved by quadratures. This, however, can be done only
after the finding special new variables which separate the associated
Hamilton-Jacobi equation. Recall, for the Holt system the additional
second integrals $K$ are the polynomials of the third, fourth and sixth
order in momenta \cite{ho82,hi87}, respectively. Therefore, it seems that
the Hamiltonians (\ref{holt}) can not be separable in the standard
curvilinear coordinate systems.

But at $\b=1,6$ the Holt system belongs to the family of the St\"{a}ckel
systems and the separation variables are related to the usual curvilinear
coordinates according to \cite{ts98b}. In fact, rescaling constant $\a$
and $\g$ in (\ref{holt})
\[
\a\to 4\,\left(\,\dfrac32\,\right)^{1/3}\,\a\,,\qquad
\g\to\dfrac{\g}{3\,\a}\]
and performing the canonical change of variables at first proposed in
\cite{hi83}
\ben
X&=&\dfrac23\,x^{3/2}\,,\qquad  p_X=p_x\,\sqrt{\,x\,}\,,\nn\\
\ll{tr1}
Y&=&-\dfrac1{2\,\sqrt{3\,\a}}\,p_y\,,\qquad p_Y=2\,\sqrt{\,3\,\a\,}\,y\,.\nn
\en
the Hamilton function (\ref{holt}) becomes
\bq
H=\dfrac{\,p_x^2+p_y^2\,}{2\,x}+2\,\a\,\,
(\,\b\,x^2+3\,y^2\,)+\dfrac{2\,\g}x\,.
\ll{ham}
\eq
Note, that at $\b=1,6$ the second additional integral of motion $K$ is a
quadratic polynomial in momenta $\{p_x,p_y\}$, what related with separability
of the Hamilton-Jacobi equation in rotated cartesian coordinates for
$(i)$ and in parabolic coordinates for $(ii)$ \cite{ts98b}.

According to \cite{ts98b}, integrals of motion for the Holt system may be
transformed by the rule
\bq
H\to\te{H}=x\,H\,,\qquad
K\to\te{K}=K-\dfrac{y^n}3\,H\,,
\qquad n=\left[\,\sqrt{\b\,}\,\right]=1,2,4\,,
\ll{hht}
\eq
into the integrals of motion for the Henon-Heiles system
\bq
\te{H}=\dfrac{\,p_x^2+p_y^2\,}{2}+2\,\a\,x\,(\,\b\,x^2+3\,y^2\,)+2\,\g
=T+V\,.
\ll{tham}
\eq
Here the ratio of the Hamiltonians $H$ and $\te{H}$ are equal to the
ratio of the determinants of the associated St\"{a}ckel matrices, which
is independent on the potential parts $V$ (\ref{tham}) of the
Hamiltonians \cite{ts98b}. On the other hand this duality may be
considered as the coupling-constant metamorphosis between integrable
systems \cite{hgdr84} with respect to the constant $\g$ in the potential
$V$.

The purpose of this letter is to show as the non-canonical transformation
(\ref{hht}) acts on the Lax matrices $L(\l)$ and $A(\l)$ in
the Lax equation
\bq
\dfrac{d\,L(\l)}{d\,t}=\{H,L(\l)\}=\left[\,A(\l),L(\l)\,\right]
\ll{lax}
\eq
and on the corresponding spectral
curve
\bq
\C(z,\l):\qquad \det(z\,I+L(\l))=0\,.\ll{sc}
\eq
The Lax representations for all the integrable cases of the Henon-Heiles
system was constructed in \cite{fo91} by using connection with stationary
flows of some known integrable PDEs. Namely these Lax pairs we shall use
to discuss the Lax representations for the Holt systems by exploiting
transformation (\ref{hht}). In \cite{rav93} the separability and another
Lax pairs for the Henon-Heiles system have been considered. We shall use
these results to construct non-canonical transformation of the
Hamiltonian (\ref{holt}) into the St\"{a}ckel form at $\b=16$ $(iii)$ .


\section{Results}
\setcounter{equation}{0}
We begin with the known Lax matrices $\te{L}(\l)$ and $\te{A}(\l)$ for
the Henon-Heiles system \cite{fo91} to construct the new Lax
representations for all the Holt system . For our present purposes it is
more convenient to use a different version of the Lax representation
obtained from that presented in \cite{fo91,brw94}. Each of the Lax
matrices $\te{L}(\l)$ will be presented as a sum of the two standard
matrices on the loop subalgebras $\L(sl(2))$ in $\L(sl(3))$ and the third
term may be associated with the outer automorphism of the whole loop
algebra \cite{ts96b}.

\vskip0.3cm
\par\noindent
{\it Case (i).}
\vskip0.3cm
Let us begin with the Lax pair for the Henon-Heiles system at $\b=1$ in
(\ref{tham}). Namely, for brevity put $\a=a^2/3$ and introduce the Lax
matrices \cite{fo91,brw94}
\ben
\te{L}(\l)&=&
\left(\begin{array}{ccc}
6\,x\l&0&9\,\l-\frac{3}{2\,a}\,p_x\\
9\,\l+\frac{3}{2\,a}\,p_x&-3\,x\l&0\\
0&9\l^3&-3\,x\l\end{array}\right)\,+
\dfrac1{\l}\,\left(\begin{array}{ccc}
0&0&0\\
0&-p_y\,y&y^2\\
0&-p_y^2&p_y\,y\end{array}\right)\nn\\
\nn\\
\nn\\
&-&a\,(3\,x^2+y^2)
\left(\begin{array}{ccc}
0&1&0\\
0&0&0\\
1&0&0\end{array}\right)\,,\qquad a=\sqrt{3\,\a\,}\,,\nn\\
\nn\\
\ll{laxi}\\
\te{A}(\l)&=&\left(\begin{array}{ccc}
0&2\,a\,\l&0\\
0&0&1\\
2\,a\,\l&-4\,a^2\,x&0\end{array}\right)\nn
\en
in the Lax equation (\ref{lax}) for the Henon-Heiles system. The spectral
curve (\ref{sc}) of the Lax matrix $\te{L}(\l)$ (\ref{laxi})
\bq
\te{\C}_1:\qquad \a\,z^3+729\,\l^7
-162\,\te{H}\,\l^3+324\,\g\,\l^3+\dfrac{\te{K}^2}{\l}=0\,,
\ll{curve1}
\eq
is a third order cyclic  covering of the line. Note, it is a very
particular case in the class of generic trigonal algebraic curves.

Now we turn to the Holt system and non-canonical transformation
(\ref{hht}). Namely, the Lax matrices for the Holt system at $\b=1$ and
$\a=a^2/3$ read as
\[
L(\l)=\te{L}(\l)+
\dfrac{3}{2\,a}\,H\,\left(\begin{array}{ccc}
0&1&0\\
0&0&0\\
1&0&0\end{array}\right)\,,\qquad
A(\l)=\dfrac{1}{x}\,\te{A}(\l)\,.
\]
The spectral curve of this matrix $L(\l)$ takes the following trigonal
form
\[
\C_1:\qquad \a\,z^3-54\,H\,z\,\l^2+729\,\l^7+324\,\g\,\l^3+\dfrac{K^2}{\l}=0\,.
\]
After point transformation
\bq
u=\dfrac12\,(x+y)\,,\qquad v=\dfrac12\,(x-y)\,,
\eq
the integrals of motion for the Henon-Heiles system became
\ben
\te{H}&=&p_u^2+p_v^2+\a\,(u^3+v^3)+2\,\g\,,\nn\\
\ll{ham1}\\
\te{K}&=&p_u^2-p_v^2+\a\,(u^3-v^3)\,.\nn
\en
The same change of the variables for the Holt systems leads to
\ben
H&=&2\,\dfrac{p_u^2+p_v^2+\a\,(u^3+v^3)+2\,\g}{u+v}\,,\nn\\
\ll{ham2}\\
K&=&2\dfrac{v\,(p_u^2+\a\,u^3+\g)-u\,(p_v^2+\a\,v^3+\g)}{u+v}\,.\nn
\en
For both system $u,v$ are separation variables and these systems belong
to the St\"{a}ckel set of the integrable systems \cite{ts98b}. In these
separation variables the Henon-Heiles dynamics splitting on two tori.
Thus, according to \cite{es96}, we can construct another $2\times 2$ Lax
representation for the Henon-Heiles system with hyperelliptic spectral
curve. Notice, the non-canonical transformation $C\to\te{C}$ (\ref{hht})
rearranges moduli $H$ and $\te{H}$ and preserves the genus of the
corresponding spectral curves. Therefore, by using a slightly different
covering of the two tori the $2\times 2$ Lax representation for the Holt
system at $\b=1$ may be constructed as well.

\vskip0.3cm
\par\noindent
{\it Case (ii).}
\vskip0.3cm
\par\noindent
Here we shall use the general construction of the Lax
representations for the dual St\"{a}ckel systems proposed in \cite{ts98b}.
For the Henon-Heiles system the Lax matrices are given by
\ben
\te{L}(\l)&=&
\left(\begin{array}{cc}p_x/2&\l-x\\
0&-p_x/2\end{array}\right)+\dfrac1{4\,\l}\,
\left(\begin{array}{cc}p_y\,y&-y^2\\
p_y^2&-p_y\,y\end{array}\right)-\nn\\
\nn\\
\nn\\
&-&6\,\a\,\Bigl[\,\l^2+x\,\l+\dfrac{4\,x^2+y^2}4\,\Bigr]\,
\left(\begin{array}{cc}0&0\\
1&0\end{array}\right)\,,\nn\\
\nn\\
\ll{laxii}\\
\te{A}(\l)&=&\left(\begin{array}{cc}0&1\\
\\
-6\,\a\,(\,\l+2\,x\,)&0\end{array}\right)\,.\nn
\en
The spectral curve (\ref{sc}) of the Lax matrix $\te{L}(\l)$ (\ref{laxii})
\bq
\te{\C}_2:\qquad z^2+6\,\a\,\l^3-\dfrac12\,\te{H}+\g+\dfrac{\te{K}}{4\,\l}=0
\ll{curve2}
\eq
is a second order covering of the line or a hyperelliptic curve.

For the Holt system the associated Lax matrices have the form
\[
L(\l)=\te{L}(\l)+\dfrac12\,H\,
\left(\begin{array}{cc}0&0\\
1&0\end{array}\right)\,,\qquad
A(\l)=\dfrac1{x}\,\te{A}(\l)\,.
\]
The corresponding spectral curve
\[
\C_2:\qquad z^2+6\,\a\,\l^3-\dfrac12\,H\,\l+\g+\dfrac{K}{4\,\l}=0
\]
remains the the same genus hyperelliptic curve.

According to \cite{ts98b}, by using the standard parabolic coordinates
\ben
u&=&\dfrac{x-\sqrt{x^2+y^2}}2\,,\qquad
p_u=p_x-\dfrac{\sqrt{x^2+y^2}+x}{y}\,p_y\,,\nn\\
v&=&\dfrac{x+\sqrt{x^2+y^2}}2\,,\qquad
p_v=p_x+\dfrac{\sqrt{x^2+y^2}-x}{y}\,p_y\,,\nn
\en
we can transform integrals of motion for the Holt and for the
Henon-Heiles systems at $\b=6$  into the St\"{a}ckel form
\ben
H&=&\dfrac{u\,(\,p_u^2+u^3\,)-v\,(\,p_v^2+v^3\,)}{u^2-v^2}\,,\nn\\
\nn\\
K&=&\dfrac{v^2\,u\,(\,p_u^2+u^3\,)-u^2\,v\,(\,p_v^2+v^3\,)}{u^2-v^2}\,,\nn\\
\nn\\
\te{H}&=&(u+v)\,H\,,\qquad \te{K}=K+\dfrac{u\,v}{u+v}\,\te{H}\,. \nn
\en
So, the Holt system at $\b=6$ belongs to the St\"{a}ckel family of
integrable systems.

\vskip0.3cm
\par\noindent
{\it Case (iii).}
\vskip0.3cm
The Lax representation  for the Henon-Heiles
system at $\b=16$ in (\ref{tham}) takes the form (see \cite{fo91,brw94})
\ben
\te{L}(\l)&=&
\left(\begin{array}{ccc}
12\,x&0&\frac38\,\a\\
9\,\l+3\,p_x&-6\,x&0\\
0&9\,\l-3\,p_x&-6\,x\end{array}\right)+
\dfrac1{4\,\l}\,\left(\begin{array}{ccc}
-2\,p_y\,y& y^2&0\\
-2\,p_y^2&0&y^2\\
0&-2\,p_y^2&2\,p_y\,y\end{array}\right)\nn\\
\nn\\
\nn\\
&+&\left(\begin{array}{ccc}
0&0&0\\
\dfrac{6\,\a\,x\,y^2}{\l}&0&0\\
\\-24\,\a\,(\,\dfrac{24\,x^2+y^2}2+\dfrac{x\,y\,p_y}{\l}\,)&
-\dfrac{6\,\a\,x\,y^2}{4\,\l}&0\end{array}\right)\,.\nn\\
\nn\\
\ll{laxiii}\\
\te{A}(\l)&=&
\left(\begin{array}{ccc}
0&1&0\\
0&0&1\\
24\,\a\,(\l-p_x)&-48\,\a\,x&0\end{array}\right)\,.\nn
\en
As above, the spectral curve of the Lax matrix (\ref{laxiii})
\bq
\te{\C}_3:\qquad 512\a\,z^3+\dfrac{9}{8}\,\l^2-\dfrac14\,\te{H}
+\dfrac{\g}2+\dfrac{\te{K}^2}{\l^2}=0
\ll{tcurv3}
\eq
is a third order cyclic covering of the line.

For the Holt system at $\b=16$ in (\ref{ham}) the Lax matrices are
equal to
\bq
L(\l)=\te{L}(\l)+
3\,H\,\left(\begin{array}{ccc}
0&0&0\\
0&0&0\\
1&0&0\end{array}\right)\,,\qquad
A(\l)=\dfrac1{x}\,\te{A}(\l)\,.
\eq
The spectral curve (\ref{sc}) of $L(\l)$ takes the following trigonal
form
\bq
\C_3:\qquad 512\,\a\,z^3-\,H\,z+\dfrac98\,\l^2+\dfrac{\g}2+\dfrac{K^2}{\l^2}=0\,.
\ll{curv3}
\eq
The second integrals of motion $\te{K}$ and $K$ in (\ref{tcurv3}-\ref{curv3})
are the square root of the non-factorable polynomial of the fourth order
in momenta. Nevertheless, both these integrals are the rational functions in
separation variables.

Recall, that according to \cite{rav93}, the separation variables for the
Henon-Heiles may be written as \cite{ts96a}
\ben
u=-&&\dfrac{\te{K}}{\a\,y^2}-\dfrac{p_y^2}{2\,\a\,y^2}+x\,,\qquad
p_u=\dfrac{p_x}2+
\dfrac{p_y}{2\,y}\,\Bigl(\,\dfrac{p_y^2}{\a\,y^2}-6\,x+
\dfrac{2\,{\te{K}}}{\a\,y^2}\,\Bigr)\,,\nn\\
\ll{hentr}\\
v=&&\dfrac{{\te{K}}}{\a\,y^2}-\dfrac{p_y^2}{2\,\a\,y^2}+x\,,\qquad
p_v=\dfrac{p_x}2+
\dfrac{p_y}{2\,y}\,\Bigl(\,\dfrac{p_y^2}{\a\,y^2}-6\,x-
\dfrac{2\,{\te{K}}}{\a\,y^2}\,\Bigr)\,.\nn
\en
Here integral of motion $K$ as yet unspecified functions of the new
variables $(x,p_x,y,p_y)$. Now we have to substitute the separation
variables (\ref{hentr}) into the definition of the second integrals
$\te{K}$ (\ref{ham1}) and to solve the resulting equation \cite{ts96a}.
Thus, substituting an explicit value of the integral $\te{K}$ into
(\ref{hentr}) we get canonical change of variables, which transforms the
Henon-Heiles integrals (\ref{ham1}) into the following form
\ben
\te{H}&=&\dfrac{p_x^2}2+\dfrac{p_y^2}2-\a\,x\,(\,16\,x^2+3\,y^2\,)+2\g\,,\nn\\
\ll{heni}\\
\te{K}^2&=&p_y^4+4\,\a\,y^3\,p_x\,p_y-12\,\a\,x\,y^2\,p_y^2
-12\,\a^2\,x^2\,y^4-2\,\a^2\,y^6\,,\nn
\en
For the Holt system we can also substitute new variables (\ref{hentr}) into
the definition of the corresponding second integrals $K$ (\ref{ham2}) and
solve the resulting equation. Thus one gets change of variables
(\ref{hentr}), which transform integrals of motion (\ref{ham2}) into the
desired form
\bq
H=\dfrac{p_x^2}{2\,x}+\dfrac{p_y^2}{2\,x}-\a\,(\,16\,x^2+3\,y^2\,)+\dfrac{2\,\g}x\,,
\qquad
K^2=\te{K}^2+\dfrac{y^4}3\,H\,,\ll{holi}
\eq
In contrast with the Henon-Heiles case, this change of the variables
is a non-canonical transformation.

So, at the third case we have a non-canonical change of variables
\[(t,x,y,p_x,p_y)\to (\te{t},u,v,p_u,p_v)\,,\]
which transforms integrals of motion (\ref{holi}) into the St\"{a}ckel
form. Of course, such transformations are not new. As an example, the
complete Kolosoff transformation \cite{kol01} connects the St\"{a}ckel
system with the Kowalewski top, which is an integrable but
non-St\"{a}ckel system. By using such transformations we can construct
the separated equations in the Lagrangian variables
$(u,\dot{u},v,\dot{v})$ and get solutions of the equations of motion in
theta-functions. Till now, in the quantum mechanics we can not construct
a counterpart of this transformation for the Holt system.

\end{document}